\documentclass{article}

\usepackage[final]{neurips_2019}

\usepackage[utf8]{inputenc} % allow utf-8 input
\usepackage[T1]{fontenc}    % use 8-bit T1 fonts
\usepackage{hyperref}       % hyperlinks
\usepackage{url}            % simple URL typesetting
\usepackage{booktabs}       % professional-quality tables
\usepackage{amsfonts}       % blackboard math symbols
\usepackage{nicefrac}       % compact symbols for 1/2, etc.
\usepackage{microtype}      % microtypography
\usepackage{amsmath}
\usepackage{graphicx}
\usepackage{float}
\usepackage{multirow}
\usepackage{placeins}
\usepackage{authblk}
\restylefloat{table}

\DeclareMathOperator*{\argmin}{argmin}

\title{Meta-SVDD: Probabilistic Meta-Learning for One-Class Classification in Cancer Histology Images}

\author[1]{Jevgenij Gamper}
\author[1]{Brandon Chan}
\author[2]{Yee Wah Tsang}
\author[2]{David Snead}
\author[1]{Nasir Rajpoot}

\affil[1]{University of Warwick, UK}
\affil[2]{University Hospital Coventry and Warwickshire, UK}

\begin{document}

\maketitle

\begin{abstract}
  To train a robust deep learning model, one usually needs a balanced set of categories in the training data. The data acquired in a medical domain, however, frequently contains an abundance of healthy patients, versus a small variety of positive, abnormal cases. Moreover, the annotation of a positive sample requires time consuming input from medical domain experts. This scenario would suggest a promise for one-class classification type approaches. In this work we propose a general one-class classification model for histology, that is meta-trained on multiple histology datasets simultaneously, and can be applied to new tasks without expensive re-training. This model could be easily used by pathology domain experts, and potentially be used for screening purposes. 
\end{abstract}

\section{Introduction}
\label{intro}

Pathology departments across most countries in the world are experiencing severe under-staffing issues \cite{metter_trends_2019, bainbridge_s_testing_2016}. With digital slide scanners becoming increasingly ubiquitous in pathology labs, it is expected that machine learning based decision support systems would substantially reduce the workload for pathology labs \cite{colling_artificial_2019}. The slide scanners are capable of producing high-resolution multi-gigapixel whole-slide images (generally of the order of $150K{\times}100K$ pixels), resulting in a wealth of pixel data that could be honed for diagnostic and prognostic purposes. The rapidly growing research community in the area of computational pathology has developed machine learning algorithms for narrow applications such as mitotic counting, cancer grading, cancer detection \cite{balkenhol_deep_2019, liu_detecting_2017, nagpal_development_2019}. These advances are important, but implementing specific but separate approaches for numerous tasks in pathology might be impractical, and in fact may be impossible due to the distribution of target categories in the population (see Figure 1). Furthermore, algorithms are frequently built on datasets that may not be representative of the population, and therefore under-perform in practice \cite{gamper_pannuke:_2019}. As such, building general-purpose algorithms, pre-trained on multiple datasets could substantially speed up the application of machine learning tools in practice \cite{hegde_similar_2019}. Furthermore, creating general algorithms that learn from few examples, would allow to easily solve very specific and narrow tasks that contain only few learning examples. 

In this paper, we propose a general one-class classification model for histology that is meta-trained on multiple datasets and can be applied to new tasks without re-training.  
We name this approach Meta-SVDD.
Past approaches for deep anomaly detection, include deterministic or variational auto-encoding, and adversarial deep generative methods \cite{schlegl_unsupervised_2017, schlegl2019f, an2015variational}. First, VAE or GAN based methods are computationally intensive, either requiring back propagation at test time, or re-training for new tasks. Second, these generative models are built with the assumption of estimating the underlying data density well, however this assumption has been put under question \cite{nalisnick_deep_2018}. Third, both autoencoding and adversarial methods solve optimisation tasks during training that are different to the downstream objectives. To address these vulnerabilities, we propose a one-class classification model that is meta-learned with the explicit loss function for one-class classification. Our novel method allows to adapt to new tasks by only observing few examples without the need for re-training as the task specific parameters inference is amortized using a neural network. The proposed method also uses out-of- as well as in-distribution examples during meta-learning optimisation. 

\begin{figure}
  \centering
  \includegraphics[width=\textwidth]{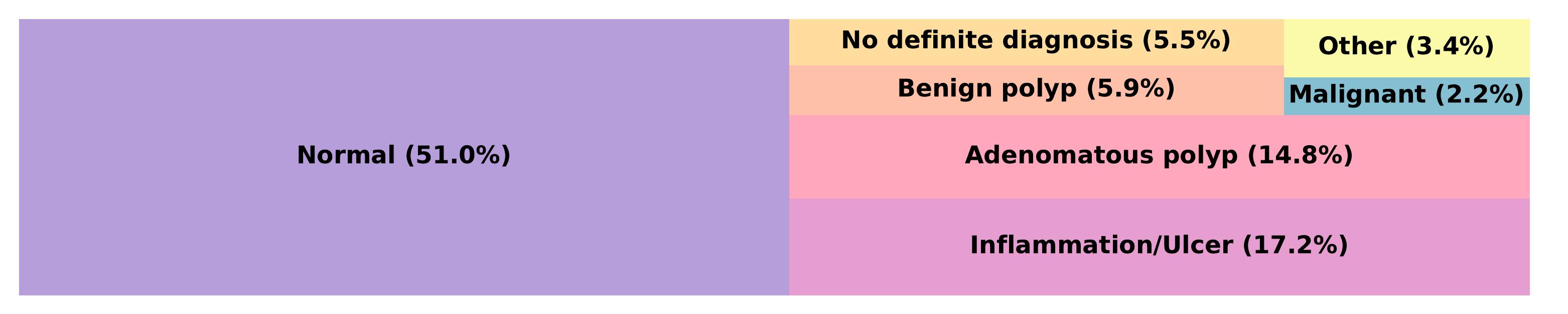}
  \caption{A breakdown by diagnosis of a sample of 23,222 patient cases from a hospital in the UK. Notice clear asymmetric distribution of categories, where for example an abnormal category has even further categorisation of various sub-abnormalities. \textit{Other} contains diagnoses of bacterial infection (Spirochete or Tuberkulosis) where only a small visual field in the whole slide image might be useful for learning.}
  \label{treemap}
\end{figure}

\section{Methods \& Results}
\label{gen_inst}

Consider a sample dataset $D^{(t)}$ for a given task $t$, which could correspond to a set of patches from histology WSIs, where $t$ corresponds to a specific tissue type (colon, lung, breast, etc.). In an ideal supervised learning task $t$ with labeled training data $D^{(t)} = \{ (\mathbf{x}_{n}^{(t)}, (\mathbf{y}_{n}^{(t)})\}^{n_t}_{n=1}$ and test data $\{ (\tilde{\mathbf{x}}_{m}^{(t)}, (\tilde{\mathbf{y}}_{m}^{(t)})\}^{M_t}_{m=1}$ pairs, the empirical marginal distribution of class labels $p(y)$ is more or less uniform \cite{buda_systematic_2018}. This allows one to estimate an optimal set of parameters $\mathbf{w}$ for a function $\mathbf{f}_{\mathbf{w}}$, i.e. a probability vector output from a convolutional neural network, via empirical risk minimization: $\argmin_{\phi} \frac{1}{N} \sum_{n=1}^{N} \mathcal{L}( \mathbf{f}_{\mathbf{w}}(\mathbf{x}_n), \mathbf{y}_n)$. Where $\mathcal{L}$ is a loss function.

\subsection{One Class Deep Support Vector Data Description (OC-SVDD)}
\label{dsvdd}

In the case, when the empirical marginal distribution of categories is not necessarily uniform, as mentioned in Section \ref{intro}, we can formulate it as a one-class classification problem: (1) reducing the complexity of the task by having only the positive, in-distribution samples, $y_i = 1, \forall y_i \in D^{(t)}$; (2) adapting the function $\mathbf{f}_{\mathbf{w}}(\mathbf{x}_n)$ to map a given sample to a latent encoding; and (3) minimizing the empirical one-class Deep SVDD (OC-SVDD) loss \cite{ruff_deep_2018}, namely
\begin{equation}
\label{equation_one}
    \mathcal{L}(\phi) = \frac{1}{N} \sum_{i=1, \forall y_i = 1} \left\|\mathbf{f}_{\mathbf{w}}(x_i) - \mathbf{c} \right\|^{2},
\end{equation}

% \begin{equation}
% \label{equation_one}
%     \argmin_{\phi} \frac{1}{N} \sum_{i=1, \forall y_i = -1} \left\|\mathbf{f}_{\mathbf{w}}(x_k) - \mathbf{c} \right\|^{2} + \frac{\eta}{N} \sum_{j=1, \forall j_i = 1} (\left\|\mathbf{f}_{\mathbf{w}}(x_j) - \mathbf{c} \right\|^{2})^{y_j},
% \end{equation}

where one learns a hyper-sphere by minimising the mean distance of all data representations to the center $\mathbf{c}$ for all positive samples \cite{ruff_deep_2018}. 

% While on the right hand side, one learns the inverse of the left hand side, pushing positive samples further from the center \cite{ruff_deep_2019}. $\eta > 0$ in Equation \ref{equation_one} is a hyper-parameter. 

\subsubsection{OC-SVDD Experiments}
\label{oc_svdd_exp}
In Table A\ref{svdd-results}, we demonstrate the results of applying OC-SVDD loss to histology data. The datasets were obtained from the following sources: Colon from \cite{kather_deep_2019}, Lung \cite{alsubaie_bottom-up_2018}, Ovary \cite{kobel_diagnosis_2010}, Lymphoma \cite{janowczyk_deep_2016}, Oral \cite{shaban_prognostic_2018}, Breast from Chamelyon Challenge\footnote{https://camelyon16.grand-challenge.org/}, and Meningioma \cite{qureshi_adaptive_2008}. 

For these experiments we took an Imagenet pre-trained ResNet18 \cite{he_deep_2015}, where we replaced the final linear layer to produce the output of dimensionality of the hyper-sphere, 128 for every task. We set batch size to 64, learning rate to $1\mathrm{e}{-4}$, and optimized over 100 epochs using ADAM \cite{kingma_adam:_2014}. The hyper-sphere center $\mathbf{c}$ is initialised using the first pass through the network. For the preliminary results presented in this paper, we did not optimise any of the hyper-parameters, these were taken from \cite{ruff_deep_2018}.

For every tissue type, we picked one of the classes and treated it as in-distribution data i.e. one task, and optimised the loss function in Equation \ref{equation_one}. For most tasks, the in-distribution data is uni-modal, and only consists of that particular category. However, in the case of Breast tissue, the category \textit{Other} contains healthy tissue, lymphocytes and other tissue phenotypes, which demonstrates the potential of OC-SVDD for tumor screening purposes.  

\subsection{Probabilistic meta-learning for SVDD (Meta-SVDD)}
\label{meta}

While the OC-SVDD method offers an explicit loss function for deep learning in one class classification and could be easily applied at test time, it still requires expensive training for any given task. One has to train 32 networks to produce the results of OC-SVDD in Table A\ref{svdd-results}. We propose to address this issue using meta-learning. We induce a distribution over function $\mathbf{f}$, by introducing an amortized distribution $q_{\phi}(\mathbf{f} | D^{(t)}, \theta)$ \cite{garnelo_neural_nodate}. We adopt the network architecture and inference method for the parameters $\phi$ according to \cite{gordon_meta-learning_2018}. Namely, during optimisation we: (i) select a task $t$ at random, (ii) sample some training data $D^{(t)}$, (iii) form the posterior predictive $q_{\phi}(\mathbf{f} | D^{(t)}, \theta)$  given in-distribution data $D^{(t)} = \{ (\mathbf{x}_{n}^{(t)}, (\mathbf{y}_{n}^{(t)})\}^{n_t}_{n=1}$, where $y_n = 1$, (iv) next we evaluate the posterior predictive on meta test data using semi-supervised SVDD loss:

\begin{equation}
\label{equation_two}
    \frac{1}{N+M+L} \left[ \sum_{i=1, \forall y_i = 1} \left\|\mathbf{f}_{l}(x_i) - \mathbf{c} \right\|^{2} + \eta \sum_{j=1, \forall y_j = -1} (\left\|\mathbf{f}_{l}(x_j) - \mathbf{c} \right\|^{2})^{y_j}\right]
\end{equation}

where $\mathbf{f}_{l} \sim q_{\phi}(\mathbf{f} | D^{(t)}, \theta)$. $L$ is the number of samples from predictive posterior. We assume $q_{\phi}$ to be Gaussian and use reparameterisation trick during optimisation \cite{kingma_auto-encoding_2013}. Compared to the Equation \ref{equation_one}, the right hand side learns the inverse of the left hand side, pushing positive samples further from the center \cite{ruff_deep_2019}, and $\eta > 0$ in Equation \ref{equation_two} is a hyper-parameter. This approach allows us to amortize the parameter learning for new tasks directly to inference network, that predicts the parameterizations for the distribution over $\mathbf{f}$ that maps test data to the latent space. We name this approach Meta-SVDD.

\subsubsection{Meta-SVDD Experiments}
\label{meta_svdd_exp}
Following \citet{gordon_meta-learning_2018}. We use the same encoder (ResNet18) for all tasks represented by $\theta$ in predictive posterior. We set $L$ to 10. We set a meta batch size to 5, and optimise using gradient accumulation, due to restrained computational resources. The inference network $q_\phi$ consists of three fully connected layers, with ReLU activation functions. The inference network takes mean of in-distribution features and produces parameterisation for posterior from which parameters of function $\mathbf{f}$ are sampled. The remaining hyper-parameters are the same as in Section \ref{oc_svdd_exp}. 

We adopt leave-one-out cross-validation setup where we pretrain on 32 tasks, and test on the remaining tasks. The results are presented in Table \ref{svdd-results}. Note that for inferring the posterior we are using only 10 in-distribution samples. Therefore the results of the proposed method are promising, however at the current stage of work we have faced the computational limitations during meta-training \cite{nichol_first-order_2018}. 

\section{Discussion \& Future Directions}
\label{results_discussion}

We present preliminary results for meta-learned one-class classification model for histology tasks, such model does not require expensive training and, parameter inference is done at test time. We demonstrated its potential for screening task in the case of Breast tissue, and flexibility with learning uni-modal tasks in other tissues. Future work would include hyper-parameter optimisation for neural network architecture, and for meta-learning. For example OC-SVDD loss resembles the tasks of self-supervised learning, and as it has been demonstrated benefits significantly from larger networks \cite{kolesnikov_revisiting_nodate}. However, that would require a careful treatment of sensitive meta-learning optimisation process \cite{antoniou_how_2018}. Once a stable set of architecture and optimisation hyper-parameters are established, we plan to thoroughly test the proposed meta learning scheme for one-class classification on whole slide images for screening and speeding up annotation. Additionally, we are planning on expanding the tasks for training using existing datasets for meta-learning \cite{triantafillou_meta-dataset:_2019}, this, which we hope would also increase the performance on fine-grained tasks such Lung adenocarcinoma subtypes. By increasing the size of the network, stabilising the optimisation process, and increasing the number of datasets, we aim to significantly improve the performance of Meta-SVDD.

\subsubsection*{Acknowledgments}

This research was partially supported by Philips Pathology.

% This research was supported by Philips Research.

% \medskip

\small
\bibliographystyle{plainnat}%{plainnat-fa}%{unsrt-fa}%{chicago-fa}{plainnat-fa}%splncs04

\bibliography{references}

\appendix

\section{Preliminary results table}

\begin{table}[!h]
\begin{tabular}{clcc} 
\hline
\multirow{2}{*}{Tissue} & \multicolumn{1}{c}{\multirow{2}{*}{Category}} & \multicolumn{2}{c}{AUC}                  \\ 
\cline{3-4}
                        & \multicolumn{1}{c}{}                          & OC-SVDD & \multicolumn{1}{l}{META-SVDD}  \\ 
\hline
Colon                   & Adipose                                       & 0.98    & 0.71                           \\ 
\cline{2-4}
                        & Background                                    & 0.94    & 0.84                           \\ 
\cline{2-4}
                        & Debris                                        & 0.77    & 0.66                           \\ 
\cline{2-4}
                        & Lymphocytes                                   & 0.94    & 0.68                           \\ 
\cline{2-4}
                        & Mucus                                         & 0.94    & 0.64                           \\ 
\cline{2-4}
                        & Muscle                                        & 0.76    & 0.67                           \\ 
\cline{2-4}
                        & Normal                                        & 0.76    & 0.69                           \\ 
\cline{2-4}
                        & Stroma                                        & 0.83    & 0.68                           \\ 
\cline{2-4}
                        & Tumor                                         & 0.80    & 0.59                           \\ 
\hline
Lung                    & Solid                                         & 0.57    & 0.52                           \\ 
\cline{2-4}
                        & Acinar                                        & 0.73    & 0.63                           \\ 
\cline{2-4}
                        & Papillary                                     & 0.75    & 0.58                           \\ 
\cline{2-4}
                        & Lepidic                                       & 0.62    & 0.57                           \\ 
\cline{2-4}
                        & Micropapillary                                & 0.67    & 0.57                           \\ 
\cline{2-4}
                        & Other                                         & 0.71    & 0.60                           \\ 
\hline
Ovary                   & High grade serous                             & 0.71    & 0.66                           \\ 
\cline{2-4}
                        & Low grade serous                              & 0.76    & 0.65                           \\ 
\cline{2-4}
                        & Endometrioid                                  & 0.60    & 0.47                           \\ 
\cline{2-4}
                        & Mucinous                                      & 0.65    & 0.59                           \\ 
\cline{2-4}
                        & Clear cell                                    & 0.52    & 0.56                           \\ 
\hline
Meningioma              & Fibr                                          & 0.85    & 0.73                           \\ 
\cline{2-4}
                        & Meningioma                                    & 0.85    & 0.74                           \\ 
\cline{2-4}
                        & Psam                                          & 0.77    & 0.72                           \\ 
\cline{2-4}
                        & Trans                                         & 0.82    & 0.73                           \\ 
\hline
Lymphoma                & Chronic lymphocytic leukemia                  & 0.71    & 0.69                           \\ 
\cline{2-4}
                        & Follicular                                    & 0.57    & 0.48                           \\ 
\cline{2-4}
                        & Mantle cell                                   & 0.60    & 0.55                           \\ 
\hline
Oral                    & Tumor                                         & 0.79    & 0.66                           \\ 
\cline{2-4}
                        & Lymphocytes                                   & 0.83    & 0.68                           \\ 
\hline
Breast                  & Tumor                                         & 0.61    & 0.57                           \\ 
\cline{2-4}
                        & Other                                         & 0.79    & 0.60                          
\end{tabular}
\caption{One-Class classification results using OC-SVDD and META-SVDD. Each of the classes in every dataset was tested using leave-one-out methodology.}
\label{svdd-results}
\end{table}
\end{document}